
%
%

\message
{JNL.TEX version 0.92 as of 6/9/87.  Report bugs and problems to Doug Eardley.}

\catcode`@=11
\expandafter\ifx\csname inp@t\endcsname\relax\let\inp@t=\input
\def\input#1 {\expandafter\ifx\csname #1IsLoaded\endcsname\relax
\inp@t#1%
\expandafter\def\csname #1IsLoaded\endcsname{(#1 was previously loaded)}
\else\message{\csname #1IsLoaded\endcsname}\fi}\fi
\catcode`@=12



\font\twelverm=cmr10 scaled 1200    \font\twelvei=cmmi10 scaled 1200
\font\twelvesy=cmsy10 scaled 1200   \font\twelveex=cmex10 scaled 1200
\font\twelvebf=cmbx10 scaled 1200   \font\twelvesl=cmsl10 scaled 1200
\font\twelvett=cmtt10 scaled 1200   \font\twelveit=cmti10 scaled 1200
\font\twelvesc=cmcsc10 scaled 1200  \font\twelvesf=amssmc10 scaled 1200
\skewchar\twelvei='177   \skewchar\twelvesy='60


\def\twelvepoint{\normalbaselineskip=12.4pt plus 0.1pt minus 0.1pt
  \abovedisplayskip 12.4pt plus 3pt minus 9pt
  \belowdisplayskip 12.4pt plus 3pt minus 9pt
  \abovedisplayshortskip 0pt plus 3pt
  \belowdisplayshortskip 7.2pt plus 3pt minus 4pt
  \smallskipamount=3.6pt plus1.2pt minus1.2pt
  \medskipamount=7.2pt plus2.4pt minus2.4pt
  \bigskipamount=14.4pt plus4.8pt minus4.8pt
  \def\rm{\fam0\twelverm}          \def\it{\fam\itfam\twelveit}%
  \def\sl{\fam\slfam\twelvesl}     \def\bf{\fam\bffam\twelvebf}%
  \def\mit{\fam 1}                 \def\cal{\fam 2}%
  \def\sc{\twelvesc}		   \def\tt{\twelvett}
  \def\sf{\twelvesf}
  \textfont0=\twelverm   \scriptfont0=\tenrm   \scriptscriptfont0=\sevenrm
  \textfont1=\twelvei    \scriptfont1=\teni    \scriptscriptfont1=\seveni
  \textfont2=\twelvesy   \scriptfont2=\tensy   \scriptscriptfont2=\sevensy
  \textfont3=\twelveex   \scriptfont3=\twelveex  \scriptscriptfont3=\twelveex
  \textfont\itfam=\twelveit
  \textfont\slfam=\twelvesl
  \textfont\bffam=\twelvebf \scriptfont\bffam=\tenbf
  \scriptscriptfont\bffam=\sevenbf
  \normalbaselines\rm}



\def\beginlinemode{\endmode
  \begingroup\parskip=0pt \obeylines\def\\{\par}\def\endmode{\par\endgroup}}
\def\beginparmode{\endmode
  \begingroup \def\endmode{\par\endgroup}}
\let\endmode=\par
{\obeylines\gdef\
{}}
\def\singlespace{\baselineskip=\normalbaselineskip}

\def\oneandahalfspace{\baselineskip=\normalbaselineskip
  \multiply\baselineskip by 3 \divide\baselineskip by 2}
\def\doublespace{\baselineskip=\normalbaselineskip \multiply\baselineskip by 2}

\newcount\firstpageno
\firstpageno=2
\footline={\ifnum\pageno<\firstpageno{\hfil}\else{\hfil\twelverm\folio\hfil}\fi}
\def\toppageno{\global\footline={\hfil}\global\headline
  ={\ifnum\pageno<\firstpageno{\hfil}\else{\hfil\twelverm\folio\hfil}\fi}}
\let\rawfootnote=\footnote		
\def\footnote#1#2{{\rm\singlespace\parindent=0pt\parskip=0pt
  \rawfootnote{#1}{#2\hfill\vrule height 0pt depth 6pt width 0pt}}}
\def\raggedcenter{\leftskip=4em plus 12em \rightskip=\leftskip
  \parindent=0pt \parfillskip=0pt \spaceskip=.3333em \xspaceskip=.5em
  \pretolerance=9999 \tolerance=9999
  \hyphenpenalty=9999 \exhyphenpenalty=9999 }
\def\dateline{\rightline{\ifcase\month\or
  January\or February\or March\or April\or May\or June\or
  July\or August\or September\or October\or November\or December\fi
  \space\number\year}}
\def\received{\vskip 3pt plus 0.2fill
 \centerline{\sl (Received\space\ifcase\month\or
  January\or February\or March\or April\or May\or June\or
  July\or August\or September\or October\or November\or December\fi
  \qquad, \number\year)}}


\hsize=6.5truein
\hoffset=0truein
\vsize=8.9truein
\voffset=0truein
\parskip=\medskipamount
\def\\{\cr}
\twelvepoint		
\doublespace		
\overfullrule=0pt	




\def\title			
  {\null\vskip 3pt plus 0.2fill
   \beginlinemode \doublespace \raggedcenter \bf}

\def\author			
  {\vskip 3pt plus 0.2fill \beginlinemode
   \singlespace \raggedcenter\sc}

\def\affil			
  {\vskip 3pt plus 0.1fill \beginlinemode
   \oneandahalfspace \raggedcenter \sl}

\def\abstract			
  {\vskip 3pt plus 0.3fill \beginparmode
   \oneandahalfspace ABSTRACT: }

\def\endtitlepage		
  {\endpage			
   \body}
\let\endtopmatter=\endtitlepage

\def\body			
  {\beginparmode}		

\def\head#1{			
  \goodbreak\vskip 0.5truein	
  {\immediate\write16{#1}
   \raggedcenter \uppercase{#1}\par}
   \nobreak\vskip 0.25truein\nobreak}

\def\beginitems{
\par\medskip\bgroup\def\i##1 {\item{##1}}\def\ii##1 {\itemitem{##1}}
\leftskip=36pt\parskip=0pt}
\def\enditems{\par\egroup}

\def\beneathrel#1\under#2{\mathrel{\mathop{#2}\limits_{#1}}}

\def\refto#1{$^{#1}$}		

\def\references			
  {\head{References}		
   \beginparmode
   \frenchspacing \parindent=0pt \leftskip=1truecm
   \parskip=8pt plus 3pt \everypar{\hangindent=\parindent}}

\def\referencesnohead   	
  {                     	
   \beginparmode
   \frenchspacing \parindent=0pt \leftskip=1truecm
   \parskip=8pt plus 3pt \everypar{\hangindent=\parindent}}

\gdef\refis#1{\item{#1.\ }}			

\gdef\journal#1, #2, #3, 1#4#5#6{		
    {\sl #1~}{\bf #2}, #3 (1#4#5#6)}		

\def\pr{\journal Phys. Rev., }

\def\prb{\journal Phys. Rev. B, }

\def\prl{\journal Phys. Rev. Lett., }

\def\np{\journal Nucl. Phys., }

\def\endreferences{\body}

\def\figurecaptions		
  {\endpage
   \beginparmode
   \head{Figure Captions}
}

\def\endpage			
  {\vfill\eject}

\def\endpaper			
  {\endmode\vfill\supereject}


\def\heading				
  {\vskip 0.5truein plus 0.1truein	
   \beginparmode \def\\{\par} \parskip=0pt \singlespace \raggedcenter}

\def\subheading				
  {\vskip 0.25truein plus 0.1truein	
   \beginlinemode \singlespace \parskip=0pt \def\\{\par}\raggedcenter}

\def\tag#1$${\eqno(#1)$$}

\def\align#1$${\eqalign{#1}$$}

\def\aligntag#1$${\gdef\tag##1\\{&(##1)\cr}\eqalignno{#1\\}$$
  \gdef\tag##1$${\eqno(##1)$$}}

\def\endaligntag{}

\def\overset #1\to#2{{\mathop{#2}\limits^{#1}}}
\def\underset#1\to#2{{\let\next=#1\mathpalette\undersetpalette#2}}
\def\undersetpalette#1#2{\vtop{\baselineskip0pt
\ialign{$\mathsurround=0pt #1\hfil##\hfil$\crcr#2\crcr\next\crcr}}}


\def\ref#1{Ref.~#1}			
\def\Ref#1{Ref.~#1}			
\def\[#1]{[\cite{#1}]}
\def\cite#1{{#1}}
\def\(#1){(\call{#1})}
\def\call#1{{#1}}
\def\taghead#1{}
\def\frac#1#2{{#1 \over #2}}

\def\12{{1\over2}}
\def\eg{{\it e.g.,\ }}

\def\ie{{\it i.e.,\ }}

\def\sla{\raise.15ex\hbox{$/$}\kern-.57em}
\def\leaderfill{\leaders\hbox to 1em{\hss.\hss}\hfill}
\def\twiddle{\lower.9ex\rlap{$\kern-.1em\scriptstyle\sim$}}
\def\bigtwiddle{\lower1.ex\rlap{$\sim$}}
\def\gtwid{\mathrel{\raise.3ex\hbox{$>$\kern-.75em\lower1ex\hbox{$\sim$}}}}
\def\ltwid{\mathrel{\raise.3ex\hbox{$<$\kern-.75em\lower1ex\hbox{$\sim$}}}}
\def\square{\kern1pt\vbox{\hrule height 1.2pt\hbox{\vrule width 1.2pt\hskip 3pt
   \vbox{\vskip 6pt}\hskip 3pt\vrule width 0.6pt}\hrule height 0.6pt}\kern1pt}
\def\tdot#1{\mathord{\mathop{#1}\limits^{\kern2pt\ldots}}}

\def\pmb#1{\setbox0=\hbox{#1}%
  \kern-.025em\copy0\kern-\wd0
  \kern  .05em\copy0\kern-\wd0
  \kern-.025em\raise.0433em\box0 }

\catcode`@=11
\newcount\r@fcount \r@fcount=0
\newcount\r@fcurr
\immediate\newwrite\reffile
\newif\ifr@ffile\r@ffilefalse
\def\w@rnwrite#1{\ifr@ffile\immediate\write\reffile{#1}\fi\message{#1}}

\def\writer@f#1>>{}
\def\referencefile{
  \r@ffiletrue\immediate\openout\reffile=\jobname.ref%
  \def\writer@f##1>>{\ifr@ffile\immediate\write\reffile%
    {\noexpand\refis{##1} = \csname r@fnum##1\endcsname = %
     \expandafter\expandafter\expandafter\strip@t\expandafter%
     \meaning\csname r@ftext\csname r@fnum##1\endcsname\endcsname}\fi}%
  \def\strip@t##1>>{}}

\def\citeall#1{\xdef#1##1{#1{\noexpand\cite{##1}}}}
\def\cite#1{\each@rg\citer@nge{#1}}	

\def\each@rg#1#2{{\let\thecsname=#1\expandafter\first@rg#2,\end,}}
\def\first@rg#1,{\thecsname{#1}\apply@rg}	
\def\apply@rg#1,{\ifx\end#1\let\next=\relax
\else,\thecsname{#1}\let\next=\apply@rg\fi\next}

\def\citer@nge#1{\citedor@nge#1-\end-}	
\def\citer@ngeat#1\end-{#1}
\def\citedor@nge#1-#2-{\ifx\end#2\r@featspace#1 
  \else\citel@@p{#1}{#2}\citer@ngeat\fi}	
\def\citel@@p#1#2{\ifnum#1>#2{\errmessage{Reference range #1-#2\space is bad.}%
    \errhelp{If you cite a series of references by the notation M-N, then M and
    N must be integers, and N must be greater than or equal to M.}}\else%
 {\count0=#1\count1=#2\advance\count1
by1\relax\expandafter\r@fcite\the\count0,%
  \loop\advance\count0 by1\relax
    \ifnum\count0<\count1,\expandafter\r@fcite\the\count0,%
  \repeat}\fi}

\def\r@featspace#1#2 {\r@fcite#1#2,}	
\def\r@fcite#1,{\ifuncit@d{#1}
    \newr@f{#1}%
    \expandafter\gdef\csname r@ftext\number\r@fcount\endcsname%
                     {\message{Reference #1 to be supplied.}%
                      \writer@f#1>>#1 to be supplied.\par}%
 \fi%
 \csname r@fnum#1\endcsname}
\def\ifuncit@d#1{\expandafter\ifx\csname r@fnum#1\endcsname\relax}%
\def\newr@f#1{\global\advance\r@fcount by1%
    \expandafter\xdef\csname r@fnum#1\endcsname{\number\r@fcount}}

\let\r@fis=\refis			
\def\refis#1#2#3\par{\ifuncit@d{#1}
   \newr@f{#1}%
   \w@rnwrite{Reference #1=\number\r@fcount\space is not cited up to now.}\fi%
  \expandafter\gdef\csname r@ftext\csname r@fnum#1\endcsname\endcsname%
  {\writer@f#1>>#2#3\par}}

\def\ignoreuncited{
   \def\refis##1##2##3\par{\ifuncit@d{##1}%
     \else\expandafter\gdef\csname r@ftext\csname
r@fnum##1\endcsname\endcsname%
     {\writer@f##1>>##2##3\par}\fi}}

\def\r@ferr{\endreferences\errmessage{I was expecting to see
\noexpand\endreferences before now;  I have inserted it here.}}
\let\r@ferences=\references
\def\references{\r@ferences\def\endmode{\r@ferr\par\endgroup}}

\let\endr@ferences=\endreferences
\def\endreferences{\r@fcurr=0
  {\loop\ifnum\r@fcurr<\r@fcount
    \advance\r@fcurr by 1\relax\expandafter\r@fis\expandafter{\number\r@fcurr}%
    \csname r@ftext\number\r@fcurr\endcsname%
  \repeat}\gdef\r@ferr{}\endr@ferences}


\let\r@fend=\endpaper\gdef\endpaper{\ifr@ffile
\immediate\write16{Cross References written on []\jobname.REF.}\fi\r@fend}

\catcode`@=12

\citeall\refto		
\citeall\ref		%
\citeall\Ref		%

\catcode`@=11
\newcount\tagnumber\tagnumber=0

\immediate\newwrite\eqnfile
\newif\if@qnfile\@qnfilefalse
\def\write@qn#1{}
\def\writenew@qn#1{}
\def\w@rnwrite#1{\write@qn{#1}\message{#1}}
\def\@rrwrite#1{\write@qn{#1}\errmessage{#1}}

\def\taghead#1{\gdef\t@ghead{#1}\global\tagnumber=0}
\def\t@ghead{}

\expandafter\def\csname @qnnum-3\endcsname
  {{\t@ghead\advance\tagnumber by -3\relax\number\tagnumber}}
\expandafter\def\csname @qnnum-2\endcsname
  {{\t@ghead\advance\tagnumber by -2\relax\number\tagnumber}}
\expandafter\def\csname @qnnum-1\endcsname
  {{\t@ghead\advance\tagnumber by -1\relax\number\tagnumber}}
\expandafter\def\csname @qnnum0\endcsname
  {\t@ghead\number\tagnumber}
\expandafter\def\csname @qnnum+1\endcsname
  {{\t@ghead\advance\tagnumber by 1\relax\number\tagnumber}}
\expandafter\def\csname @qnnum+2\endcsname
  {{\t@ghead\advance\tagnumber by 2\relax\number\tagnumber}}
\expandafter\def\csname @qnnum+3\endcsname
  {{\t@ghead\advance\tagnumber by 3\relax\number\tagnumber}}

\def\equationfile{%
  \@qnfiletrue\immediate\openout\eqnfile=\jobname.eqn%
  \def\write@qn##1{\if@qnfile\immediate\write\eqnfile{##1}\fi}
  \def\writenew@qn##1{\if@qnfile\immediate\write\eqnfile
    {\noexpand\tag{##1} = (\t@ghead\number\tagnumber)}\fi}
}

\def\callall#1{\xdef#1##1{#1{\noexpand\call{##1}}}}
\def\call#1{\each@rg\callr@nge{#1}}

\def\each@rg#1#2{{\let\thecsname=#1\expandafter\first@rg#2,\end,}}
\def\first@rg#1,{\thecsname{#1}\apply@rg}
\def\apply@rg#1,{\ifx\end#1\let\next=\relax%
\else,\thecsname{#1}\let\next=\apply@rg\fi\next}

\def\callr@nge#1{\calldor@nge#1-\end-}
\def\callr@ngeat#1\end-{#1}
\def\calldor@nge#1-#2-{\ifx\end#2\@qneatspace#1 %
  \else\calll@@p{#1}{#2}\callr@ngeat\fi}
\def\calll@@p#1#2{\ifnum#1>#2{\@rrwrite{Equation range #1-#2\space is bad.}
\errhelp{If you call a series of equations by the notation M-N, then M and
N must be integers, and N must be greater than or equal to M.}}\else%
 {\count0=#1\count1=#2\advance\count1
by1\relax\expandafter\@qncall\the\count0,%
  \loop\advance\count0 by1\relax%
    \ifnum\count0<\count1,\expandafter\@qncall\the\count0,%
  \repeat}\fi}

\def\@qneatspace#1#2 {\@qncall#1#2,}
\def\@qncall#1,{\ifunc@lled{#1}{\def\next{#1}\ifx\next\empty\else
  \w@rnwrite{Equation number \noexpand\(>>#1<<) has not been defined yet.}
  >>#1<<\fi}\else\csname @qnnum#1\endcsname\fi}

\let\eqnono=\eqno
\def\eqno(#1){\tag#1}
\def\tag#1$${\eqnono(\displayt@g#1 )$$}

\def\aligntag#1\endaligntag
  $${\gdef\tag##1\\{&(##1 )\cr}\eqalignno{#1\\}$$
  \gdef\tag##1$${\eqnono(\displayt@g##1 )$$}}

\def\eqalignno#1{\displ@y \tabskip\centering
  \halign to\displaywidth{\hfil$\displaystyle{##}$\tabskip\z@skip
    &$\displaystyle{{}##}$\hfil\tabskip\centering
    &\llap{$\displayt@gpar##$}\tabskip\z@skip\crcr
    #1\crcr}}

\def\displayt@gpar(#1){(\displayt@g#1 )}

\def\displayt@g#1 {\rm\ifunc@lled{#1}\global\advance\tagnumber by1
        {\def\next{#1}\ifx\next\empty\else\expandafter
        \xdef\csname @qnnum#1\endcsname{\t@ghead\number\tagnumber}\fi}%
  \writenew@qn{#1}\t@ghead\number\tagnumber\else
        {\edef\next{\t@ghead\number\tagnumber}%
        \expandafter\ifx\csname @qnnum#1\endcsname\next\else
        \w@rnwrite{Equation \noexpand\tag{#1} is a duplicate number.}\fi}%
  \csname @qnnum#1\endcsname\fi}

\def\ifunc@lled#1{\expandafter\ifx\csname @qnnum#1\endcsname\relax}

\let\@qnend=\end\gdef\end{\if@qnfile
\immediate\write16{Equation numbers written on []\jobname.EQN.}\fi\@qnend}

\catcode`@=12

\def\eg{{\it e.g.,\ }}
\def\ie{{\it i.e.,\ }}

\def\>{\rangle}
\def\<{\langle}
\def\o{\over}

\def\slD{\raise.15ex\hbox{$/$}\kern-.57em\hbox{$D$}}
\def\dsl{\raise.15ex\hbox{$/$}\kern-.57em\hbox{$\Delta$}}
\def\slp{{\raise.15ex\hbox{$/$}\kern-.57em\hbox{$\partial$}}}
\def\nsl{\raise.15ex\hbox{$/$}\kern-.57em\hbox{$\nabla$}}
\def\sla{\raise.15ex\hbox{$/$}\kern-.57em\hbox{$\rightarrow$}}
\def\slla{\raise.15ex\hbox{$/$}\kern-.57em\hbox{$\lambda$}}
\def\slb{\raise.15ex\hbox{$/$}\kern-.57em\hbox{$b$}}
\def\lnp{\raise.15ex\hbox{$/$}\kern-.57em\hbox{$p$}}
\def\lnk{\raise.15ex\hbox{$/$}\kern-.57em\hbox{$k$}}
\def\lnK{\raise.15ex\hbox{$/$}\kern-.57em\hbox{$K$}}
\def\lnq{\raise.15ex\hbox{$/$}\kern-.57em\hbox{$q$}}

\def\a{\alpha}

\def\ga{{\gamma}}
\def\de{{\delta}}
\def\eps{{\epsilon}}

\def\la{\lambda}

\def\Om{{\Omega}}

\def\cJ{{\cal J}}

\def\cL{{\cal L}}

\def\part{\partial}

\def\dag{\dagger}

\def\abs{
         \vskip 3pt plus 0.3fill\beginparmode
         \doublespace ABSTRACT:\ }

\singlespace
\def\skiip#1{}

\title
Orbital spins of the collective excitations in Hall liquids

\author Dung-Hai Lee

\affil
IBM Research Division,
T.J. Watson Research Center,
Yorktown Heights, NY 10598

\author Xiao-Gang Wen

\affil
Department of Physics
MIT
77 Massachusetts Avenue
Cambridge, MA 02139

\abs{
We show that the Chern-Simons-Ginzburg-Landau theory of the quantum
Hall effect needs a modification, because the order parameter in that
theory carries an intrinsic orbital angular momentum.
This quantum number contains additional information about the topological
order of the Hall liquids. We propose to measure this angular momentum
by circular-polarized Ramman scattering.
}

\endtopmatter

It is now well known that the states of a two-dimensional
electron gas that exhibit quantized Hall conductivity $\sigma_{xy}$
are incompressible quantum liquids (the Hall liquids). It turns out that
$\sigma_{xy}$ alone is {\it insufficient} to specify a Hall liquid.
In order to do that one also needs to know the hierarchy
scheme\refto{HH,Hi} by which the Hall liquid is constructed.
Loosely speaking, a hierarchy is specified by a matrix\refto{Mtx}
and $\sigma_{xy}$ is only the (1,1) element of its inverse.
The purpose of this paper is to answer the question
``what kind of experiments bare information about the rest of the
matrix''. We demonstrate that a quantum number called ``orbital spin''
\refto{WZ} carried by all elementary excitations of a Hall liquid
contains such information. In particular, we purpose to measure
the orbital spins of the collective excitations by
circular-polarized Ramman scattering.

Before going into details, we first summarize our main results.
i) We show that the Chern-Simons-Ginzburg-Landau theory\refto{ZHK}
of the quantum Hall effect needs a modification because the order
parameter\refto{GM}
in that theory carries an intrinsic orbital angular momentum.
This angular momentum (the orbital spin) is constrained to be
locally perpendicular to the two-dimensional surface. There are
some physical consequences due to the existence of the orbital spin. Some of
them (e.g. the Berry phase\refto{Berry} caused by the motion
of orbital spins) are only effective on {\it curved} two dimensional
surfaces, and others (e.g. the selection rule) are applicable on
{\it flat} surfaces as well. ii) We calculate the values of the orbital
spin carried by the elementary excitations of Hall
liquids, and use those to explain features in the numerical spectra.
To measure the orbital spin, iii) we propose and predict the outcome
of circular polarized Ramman scattering using the orbital spin selection rule.

To get a physical feeling for the orbital spin, let us think of the motion
of an electron in magnetic field as the superposition of the fast
cyclotron and slow guiding-center motions. If we concentrate on the
slow degrees of freedom, we may think of the cyclotron motion as giving
rise to an orbital spin to the guiding center.
The value of the orbital spin is determined by the Landau level index
of the cyclotron motion.

It turns out that the orbital spin has direct consequences on the
spectra of Hall liquids. In Fig. 1 we present the excitation spectra
of the $\nu=1/3$ and $\nu=2/5$ Hall liquids.
These results are obtained by diagonalizing the Hamiltonian describing
a finite number of electrons on a sphere with Coulomb interaction.
A careful examination of Fig.1 reveals several
differences between these two cases.
a) The minimum angular momenta of the
collective mode are $2$ in the $2/5$ liquid, and $1$ in the $1/3$ liquid.
b) The collective mode of the $1/3$ liquid
merges with the continuum at small angular momenta,
while that associated with the $2/5$ liquid remains
separated from the continuum.
c) At small angular
momenta, the $\nu=1/3$ collective mode disperses downward while
the $\nu=2/5$ mode disperses
upward.\refto{SW}

First we discuss the origin of orbital spins.
Following \Ref{ZHK} we consider the following
Chern-Simons-Ginzburg-Landau (CSGL) action for an $\nu$ (the filling
factor) $=1/m$ ($m=$ an odd integer) Hall liquid.
$$\eqalign{
\cL &=\Phi^\dag(\part_0-iA_0+ia_0)\Phi
-{1\over 2m_e}\Phi^\dag(\part_j-iA_j+ia_j)^2\Phi
+{1\over 2}(\Phi^\dag\Phi-\rho)V(\Phi^\dag\Phi-\rho) \cr
&-{i\over 4\pi m}\eps a\part a\cr}
\eqno(SGL)$$
where $\Phi$ is the Girvin-MacDonald order parameter (and hence is a
Bose field),
$V$ a two-body potential (including a
hare-core contribution), $\rho$ the average electron density,
$A_\mu$ the external electromagnetic gauge field, $a_\mu$ the
statistical gauge field.
In \(SGL) and the rest of the paper $\eps\alpha\part\beta=\eps^{\mu\nu\la
}\alpha_\mu\part_\nu\beta_\la$.
The last term in \(SGL), the Chern-Simons term,
ensures that $m$ flux quanta of the statistical gauge field are attached
to each $\Phi$-boson (the condensate boson) and hence transmutes
it back into the original electron.
Let $J_\mu=\sum_k j_{k\mu}$ (where
$j_{k0}=\delta(\vec{x}-\vec{x}_k)$, and
$\vec j_{k}=\dot {\vec x}_{k} \delta(\vec{x}-\vec{x}_k)$),
be the Bose 3-current, the term
$$
\cL_{link}={i\over 2}(2\pi m)\sum_{k\ne k^\prime}\eps j_k {1\over \part^2
}\part j_{k^\prime}+
{i\over 2}(2\pi m)\sum_k\eps j_k {1\over \part^2
}\part j_k
\eqno(Slink1)$$
is generated after integrating out $a_\mu$
under the gauge $\part\cdot a= \part^\mu a_\mu =0$.
In \(Slink1) ${1\over \part^2} ={1\over \part^\mu\part_\mu}$.
Eq. \(Slink1) contains two terms. The first, mutual-interaction, term
gives a phase factor
$exp(-im\pi)$ whenever two bosons braid around each other.
As the result, the statistics of the $\Phi$-boson is transmuted back
to that of a fermion.
The second, self-interaction, term causes the coupling of the
electronic motion to curvature
upon using the following identity\refto{Pkv}
$$
2\pi\eps j_k {1\over \part^2}\part j_k =j_k\cdot\omega
\eqno(Slink2)$$
In the above $\omega_0=0$ and $\epsilon_{ij}\part_i\omega_j$
is the curvature of space.\refto{WZ} In order to remove this
self-interaction effect, a counter term
must be added to \(SGL) so as to cancel the second term
in Eq.(2).

The new CSGL action which includes this counter term is given by
$$\eqalign{
\cL &=\Phi^\dag(\part_0-iA_0+ia_0-i {m\over 2}\omega_0)\Phi
-{1\over 2m_e}\Phi^\dag(\part_j-iA_j+ia_j-i{m\over 2}
\omega_j)^2\Phi \cr
&+{1\over 2}(\Phi^\dag\Phi-\rho)V(\Phi^\dag\Phi-\rho)
-{i\over 4\pi m}\eps a\part a\cr}
\eqno(SGL1)$$
Since the (two-dimensional) space curvature is given by
$\part_1\omega_2-\part_2\omega_1$, in flat space
$\omega_\mu$ is a trivial gauge and has no effect.
In a curved space (e.g. a sphere), local curvature simulates the effect
of external magnetic flux density and cause shift in the magnetic flux
density.

A straightforward generalization of the
duality transformation\refto{LF} to allow for the additional flux due
to the space curvature gives the following effective
theory for the $\nu=1/m$ quantum Hall liquid in curved space\refto{WZ}.
$$\cL=
{i\over 4\pi}
[ m\eps b \part b -\eps
(2A+m\omega)\part b
]
-i\cJ\cdot b
\eqno(Seff)$$
In the above
$b_\mu$ is related to the particle 3-current
$J_\mu$ via $J^\mu={1\o 2\pi} \eps^{\mu\nu\la}\part_\nu b_\la$,
and $\cJ_\mu$ is the vortex 3-current.
If the space is a 2-sphere one can show that by setting
$\part\cL/\part b_0 =0$, when
$N_\phi+m=mN_e$ ( where $N_e$ is the
total particle number $N_e$, and $N_{\phi}$ is the
total magnetic flux-quantum number) $\cJ_\mu=0$ and hence the
system is incompressible. Therefore as a result of the
coupling to the space curvature, a shift\refto{WZ} $\Delta N_\phi=m$
in $N_\phi$ for the appearance of a $\nu=1/m$ Hall liquid is
resulted.\refto{WZ}

Using the \(Seff) we can also
determine orbital-spins of the
the quasiparticles and quasiholes.
By integrating out $b_\mu$ in \(Seff) under the gauge $\part\cdot b
=0$
we obtain the following
quasiparticle-quasihole effective action
$$\cL=-{i\over 4\pi m}\eps A\part A
-{i\over 2m}(2\pi) \eps\cJ {1\over \part^2} \part\cJ
-{i\over m} \cJ\cdot A
-{i\over 2} \cJ\cdot\omega
\eqno(Sqp)$$
By substituting
$\cJ_\mu=q j_\mu$ (where $j_\mu\equiv \left(
\de(\vec x -\vec x_0),\dot{\vec x}
\de(\vec x -\vec x_0)
\right)$)
into \(Sqp) and use \(Slink2)
we obtain the quasiparticle ($q=1$) and quasihole ($q=-1$) orbital-
spins as
$S_{qp}={1\over 2}+{1\over 2m}$,
and
$S_{qh}=-{1\over 2}+{1\over 2m}$.

When $m=1$ (\ie $\nu=1$) $S_{qp}=1$ and
$S_{qh}=0$. Therefore, a quasihole may not be viewed as
the antiparticle of a quasielectron.
This is due to the fact that while
a quasihole is a missing electron in the {\it first} Landau level,
a quasielectron is an excess electron in the {\it second} Landau level.
The extra orbital spin carried by the quasielectrons reflects the
fact that the guiding-center basis for the second Landau level are
that of the first Landau level multiplied by $\bar{z}-\bar{z}_g$ (where
$z_g$ is the complex coordinate of the guiding-center).
Similarly, we can calculate the minimum orbital spin associated with
a pair of quasielectrons (quasiholes) by
letting $\cJ_\mu= 2j_\mu$ ($\cJ_\mu= -2j_\mu$) in \(Sqp). The result
is $S'_{qp}= 1+{2\o m}= 3$ and $S'_{qh}=-1+{2\o m}=1$.
The fact that $S'_{qp,qh}\ne 2S_{qp,qh}$ reflects the fact that when two
quasielectrons (quasiholes)
are put into the second (first) Landau level the Pauli exclusion principle
forces them to have a minimum relative angular momentum 1($-1$).
Therefore after adding
the orbital spin quantum number, the effective theory is enriched to
contain the Landau level structure and exclusion principle.

We now consider a $\nu=2/(2m-1)$ ($m$ is odd) Hall liquid.
Let us first
calculate $\Delta N_\phi$.
As a second-level hierarchical state, the
$\nu=2/(2m-1)$ Hall liquid is a Laughlin-condensate
of the $1/m$ quasiparticles on top of the $1/m$ QH state.
After adding the appropriate counter terms
at both levels of the hierarchy, it is straightforward
to generalize the derivation of \(Seff) to obtain\refto{WZ}
$$\cL={i\over 4\pi} [
K_{n^\prime n} \eps b_{n^\prime }
\part b_n - \eps (2t_n A+2S_n\omega)
\part b_n
]
- i\cJ_n\cdot b_n
\eqno(Seff1)$$
where $n$ is the level index, and $K_{n^\prime n}$, $t_n$, and $S_n$ are given
by
$K=\pmatrix{m,& -1\cr -1,& 2\cr}$,
$t=\pmatrix{1\cr 0\cr}$ and $S=\pmatrix{m/2\cr 1\cr }$.
By setting $\part\cL/\part b_{n0}=0$, we obtain
$(\cJ_n)_\mu=0$ and hence an incompressible ground state
when $N_\phi+m+1=(m-{1\over 2})N_e$. Therefore
$\Delta N_\phi=m+1$.

We now calculate the orbital-spins of quasiparticle and quasihole.
By integrating out $(b_n)_\mu$ in \(Seff1)
under the gauge $\part\cdot b_n=0$, we
obtain the following quasiparticle-quasihole effective action
$$
\cL=-{i\over 4\pi}{2\over 2m-1} \eps A \part A - it^\prime_n\cJ_n\cdot A
-iS^\prime_n\cJ_n\cdot\omega - i(2\pi)
K^\prime_{mn}\eps\cJ_m {1\over \part^2}\part\cJ_n
\eqno(Sqp1)$$
In \(Sqp1) $K^\prime_{mn}$, $t^\prime_n$, and $S^\prime_n$ are given by
$K^\prime=K^{-1}={1\o 2m-1} \pmatrix{2,& 1\cr 1, & m\cr}$,
$t'={1\o 2m-1}\pmatrix{2\cr 1\cr}$ and
$S'={1\o 2m-1}\pmatrix{m+1\cr 3m/2 \cr}$.
Using \(Slink2) we calculate the orbital-spins of
the quasiparticle/quasihole created by letting
$\left( (\cJ_1)_\mu, (\cJ_2)_\mu\right) =(q j_\mu ,0)$
(with $q=\pm 1$), and the result is
$S_{1,qp}={m+2\o 2m-1}$ and $S_{1,qh}=-{m\o 2m-1}$. Similarly we obtain
the orbital-spins of
the quasiparticle/quasihole created by letting
$\left( (\cJ_1)_\mu, (\cJ_2)_\mu\right) =(0, q j_\mu)$
to be $S_{2,qp}={2m\o 2m-1}$ and $S_{2,qh}=-{m\o 2m-1}$.
These two types of quasihole/quasiparticle are induced by inserting
an unit vortex/antivortex in the first/second-level condensate
respectively.

The collective excitations in the
$\nu=2/(2m-1)$ Hall liquid are in general made up of
$q_1$ level-1 and $q_2$ level-2 vortices respectively.
Unlike in a primary Hall liquid where spatial-coinciding vortex
and antivortex annihilate, in a hierarchy Hall liquid vortex and
antivortex from different levels can overlap without annihilating.
In particular, by letting
$\left( (\cJ_1)_\mu, (\cJ_2)_\mu\right) =(q_1 j_\mu ,q_2 j_\mu)$
(therefore $q_1$ level-1 and $q_2$
level-2 vortices are sitting on top of each other), and using \(Slink2),
we obtain the following charge and the orbital-spin
$$\eqalign{
Q(q_1,q_2) &={1\over 2m-1}(2q_1+q_2) \cr
S(q_1,q_2) &={1\o 2m-1}[(m+1)q_1+3mq_2+q_1^2+mq_2^2+q_1q_2]\cr}
\eqno(QS)$$
To understand the physical meaning of the
different excitations labeled by
$(q_1,q_2)$, let us consider the case where $m=1$ (i.e. $\nu=2$).
The excitations labeled by $(q_1,q_2)=(0,-1)$ and $(-1,1)$ carry charge
$Q=-1$ and orbital-spin $S=-1$ and $0$, therefore they
correspond to a hole in the second and the first Landau level
respectively. Similarly, the excitation $(q_1,q_2)=(0,1)$ carries
charge $Q=1$ and orbital-spin $S=2$, hence corresponds to an electron
in the third Landau level. Finally, the excitation $(q_1,q_2)=(1,-1)$
with $Q=1$ and $S=1$ has the quantum numbers of an electron in the second
Landau level. However due to the Pauli principle, such an excitation
is not allowed since in the ground state the second Landau level
is already full. This discussion taught us an important lesson, namely,
not all $(q_1,q_2)$ correspond to allowed excitations. Indeed,
when using the effective theory to
describe excitations in a Hall liquid, we have to add a
supplementary
rule that
when an excitation contains vortices that have
already condensed, it should be disallowed
(due, in integer QH effect, to the Pauli exclusion
principle, and, in fractional QH effect, to energetic considerations).
We need to keep this point close in mind when
using the effective theory to describe excitations in Hall liquids.

For general $m$,
among the various $(q_1,q_2)$ pairs, neutral excitations (Q=0) arise
when $q_2=-2q_1$. The orbital spin associated with these
$(q_1,q_2)=(n,-2n)$ excitation is $S=n(n-1)$.
Among the allowed excitations (i.e. $n<0$ according to the
supplementary rule), the excitation $(q_1,q_2)=(-1,2)=(-1,1)+(0,1)$
carries the minimum orbital-spin $S=2$. In the $m=1$ (i.e. $\nu=2$)
case, this excitation corresponds to a hole in the first and an
electron in the third Landau level. In the $m=3$ (i.e. $\nu=2/5$) case,
we identify it with the collective mode in the gap of the
$\nu=2/5$ spectrum. Finally, in the effective theory
the low-momentum cyclotron mode is described by the Gaussian fluctuation in
the $b_{1,\mu}$ gauge field.

Since the ($-$1,2) excitation can be viewed as an overlapping ($-$1,1)
quasihole and (0,1) quasiparticle, a natural question is
what happen when we pull the ($-$1,1) and (0,1) vortices apart.
Let $\vec{d}$ be the relative displacement,
it is simple to show that for small
$\mid\vec{d}\mid$ the following change in the Lagrangian is induced
$$\Delta\cL = {1\over 2} K \mid\vec{d}\mid^2 + iQ H_0\eps_{ij}d_i
\dot{R}_j
\eqno(DL)$$
In \(DL) $\vec{R}$ is the position vector associated with the
center-of-mass of the dipole,
$Q={1\over 2m-1}$ is the charge of the (0,1) quasiparticle,
$H_0$ is the strength of the external magnetic flux density, and
$K$ is an effective spring constant.
The first term in \(DL) describes the change in Coulomb energy
caused by the relative displacement, and the second term is the
associated Aharonov-Bohm phase. To calculate the partition function
we integrate over both $\vec{R}$ and $\vec{d}$. After $\vec{d}$ is
integrated over a term of the form
$\cL = {1\over 2}{(QH_0)^2\over K}\mid\dot{\vec{R}}\mid^2$
is generated, as the result the total action describing the
center-of-mass motion of the dipole is
$$\cL={M\over 2}
\mid\dot{\vec{R}}\mid^2 - iS\dot{\vec{R}}\cdot\vec{\omega}
\eqno(CM)$$
where $M=(QH_0)^2/K$, and $S=2$ is the orbital spin of the
($-$1,2) excitation. The second term describes the coupling between the
orbital-spin and the curvature.
On a sphere where $\int d^2x (\part_1\omega_2-\part_2\omega_1)=2$,
\(CM) describes a particle with effective
mass $M$ and charge S(=2) moving under the influence of a
monopole carrying two Dirac flux quanta sitting at the spherical
center. The effective magnetic field seen by the particle is
$H_{eff}={1\o R^2}$. The
spectrum associated with \(CM) has $E={1\over 2}\hbar\omega_c (
L(L+1)-4)/2$, where $L\ge S=2$ and $\omega_c ={2H_{eff}\over
M}$ is the effective cyclotron frequency.
This behavior is consistent with the low-$L$
behavior of the collective mode shown in Fig.1b.
The maximum angular momentum of the ($-$1,2) branch
is realized when
the ($-$1,1) quasihole and the (0,1) quasiparticle are diametrically
opposite to each other. In that case the total Berry's phase seen
by this largest dipole upon rotation is given by
${2\over 2m-1}N_\phi+2(S(0,1)-S(1,1))={2\over 2m-1}\left({2m-1\over 2}
N_e -(m+1)\right)+2\left({2m\over 2m-1}-{m-1\over 2m-1}\right)=N_e$.
As the result, $L_{max}={N_e\over 2}$.  Therefore
if this mode were to persist for the entire range of angular momenta,
it should end at $L={N_e\over 2}$.
The fact that
$L_{max}={N_e\over 2}+1$ in Fig.1b indicates that near $L_{max}$
the collective mode is derived from
the (0,1) quasiparticle and $(0,-1)$ quasihole pair, whose
maximum angular momentum is precisely ${N_e\over 2}+1$.
Combining these facts, we conclude that
at sufficiently large angular momenta, the $(-1,2)$ mode crosses
with the collective
mode made up of the level-2 vortex-antivortex multipole,\refto{LZ}
at which point the energy
disperse downward as $L$ further increases and eventually reaches
the roton minimum.

Similar calculations can be done for the $\nu=2/7$ QH
liquid. To summarize the results we find $\Delta N_\phi=2$,
$S_{1,qp}=3/7$, $S_{1,qh}=-1/7$, $S_{2,qp}=3/7$, $S_{2,qh}=
-6/7$. The neutral excitations made up of spatial coinciding vortices
have $(q_1 ,q_2 )=(n,-2n)$ with the
associated orbital spin given by $S=-n(n+1)$. Given the fact the
$\nu=2/7$ Hall liquid is a condensate of the $\nu=1/3$ quasiholes,
the supplementary rule implies that the allowed excitations
corresponds to $n>0$. Among them, the minimum
orbital-spin excitation has $(q_1 ,q_2)=(1,-2)$ and $S=-2$.
In principle,
the $(1,-2)$ excitation, like the $(-1,2)$ excitation in the $\nu=2/5$
liquid, is allowed. However, given the fact that the 1/3-quasiparticle
usually has much higher creation energy than the 1/3-quasihole, and the
fact that the $(1,-2)$ excitation contains a
1/3-quasiparticle, we suspect it to have an excitation energy
larger than twice the roton gap, and hence lies inside the continuum.
Indeed, numerical spectrum of the 2/7 liquid shows no traces of
this excitation in the gap.

The orbital spin of the
$(-1,2)$ collective mode in the $2/5$ state can in principle be
detected by circular polarized Ramman scattering. In this experiment under
proper orientation of the magnetic field, a
right-hand polarized light beam
incident normal to the 2D electron gas
excites the $(-1,2)$ collective mode. As the result,
the polarization of photon
changes
to left handed
for forward scattering, and right handed for backward scattering.
To be specific, the differential probability $dP$ of an incident
photon gets scattered into a solid angle $d\Om$ is given by
$$dP=\ga^2 \a^2 (l_B k)^2
|(\eps_{in,x}+i\eps_{in,y})(\eps_{out,x}^{*}+i\eps_{out,y}^{*})|^2 d\Om
\eqno(om)$$
where $k$ is the norm of the wave vector of light,
$\vec \eps_{in}$
($\vec \eps_{out}$) the
polarization vector of the
incoming (outgoing) beam, $\a =1/137$ the fine structure constant,
and $\ga$ a dimensionless constant of order unity.
In deriving \(om) we have assumed
that the energy of the photon is much larger then the excitation energy
of the collective mode.

The $\nu=3/5$ state also has a well defined collective mode near
$k=0$ due to the
particle-hole symmetry. However the orbital spin of this mode is opposite to
that of the $2/5$ mode, as one can show from direct calculation or from
particle-hole conjugation. The $3/5$ mode will have an opposite effect
on the Ramman scattering, \eg for the same orientation of the magnetic field,
the 3/5 mode will change the left-hand polarized light to the right-hand one
in the forward scattering.

Finally, an interesting lesson about Jain's\refto {Jain}
mapping from the integer to the fractional
QH effect can be learned from the present work.
According to this mapping, the ground state wave function at
$\nu=2/5$ is equal to that at $\nu=2$ operated upon by the Landau-level
projection operator.
If one generalizes this idea to include the excited states, one
finds the following puzzling situation.
Under such mapping, the $(-1,2)$ collective mode
corresponds to the first to third (1-3) Landau
level particle-hole excitation of the $\nu=2$ liquid. From the
Landau level point of view it is very unnatural to see the
1-3 (instead of 2-3 ) excitation to have the lowest energy.
The resolution of this
puzzle comes with the realization of the effects of Landau-level
projection. The projection, in addition to entirely annihilates
the 2-3 excitation at $L=1$\refto{HLR}, modifies its
properties qualitatively at small angular momenta.
This constitutes an example
in which the straightforward correspondence between the
integer and fractional quantum Hall effect is spoiled by the Landau
level projection.

XGW is supported by NSF grant No. DMR-91-14553

\references

\refis{HH}
F.D.M. Haldane, \prl 51, 605, 1983;
B. I. Halperin, \prl 52, 1583, 1984.

\refis{Hi}
S. Girvin, \prb 29, 6012, 1984; A.H. MacDonald and D.B. Murray,
 \prb 32, 2707, 1985; D-H Lee and M.P.A. Fisher , \prl 63, 903,
1989; J.K. Jain, \prl 63, 199, 1989; \pr B41, 7653, 1991.

\refis{Mtx}
B. Blok and X.G. Wen, \pr B42, 8133, 1990; \pr B42, 8145, 1990;
N. Read, \prl 65, 1502, 1990;
J. Fr\"ohlich and A. Zee, \np  B364, 517, 1991;
Z.F. Ezawa and A. Iwazaki, \pr B43, 2637, 1991;
X.G. Wen and A. Zee, \prb 46, 2290, 1992.

\refis{WZ}
X.G. Wen and A. Zee, \prl 69, 953, 1992.

\refis{HLR}
G. Dev and J.K. Jain, preprint; A qualitative
similar conclusion was also reached in
B. I. Halperin, P.A. Lee, and N. Read, preprint;
A. Lopez and E. Fradkin, preprint.

\refis{LZ}
D-H Lee and S-C Zhang, \prl 66, 1220, 1991.

\refis{Jain}
J.K. Jain, in \Ref{Hi}.

\refis{ZHK}
S. C. Zhang, T. H. Hansson and S. Kivelson, \prl 62, 82, 1989.

\refis{Pkv}
A.M. Polyakov, Mod. Phys. Lett. A3, 325, 1988;
F. Wilczek and A. Zee,  \prl 51, 2250, 1983.

\refis{GM}
S.M. Girvin and A.H. MacDonald,  \prl 58, 1252, 1987.

\refis{Berry}
M.V. Berry, Proc. Roy. Soc. London, Ser. A 392, 45 (1984).

\refis{LF}
D-H Lee and M.P.A. Fisher, Int. J. Mod. Phys. B 5, 2675, 1991.

\refis{SW}
The spectrum of the $2/5$ state on {\it torus} was obtained
by S.P. Su and F.Y. Wu, \pr, B36, 7565, 1987. Their results
give a collective mode dispersion that
shows double minima and disperses downward at small
momenta. We don't understand the origin of the spectral difference
between sphere and torus at present.

\endreferences


\head{Figure captions}

\item{Fig. 1} The low energy spectrums of (1a) $N_e=7$ electrons
at $\nu=1/3$ (\ie $N_\phi=18$), and
(1b) $N_e=10$ electrons at $\nu=2/5$ (\ie $N_\phi=21$) on a sphere
(first 17 states).
The interparticle interaction is Coulombic and the
energy is measured in units of ${e^2\o \eps l_B}$.

\vfil\eject

\end